\documentclass[aps,prl,10pt,amsmath,floats,floatfix,twocolumn,superscriptaddress,altaffilletter,nofootinbib,shownopacs,longbibliography]{revtex4-1}
\newcommand\prlsec[1]{\vspace{2mm}\noindent \textbf{\emph{#1}}\,---}
\usepackage{placeins}
\usepackage{subfig}
\usepackage{natbib}
\usepackage{orcidlink}
\usepackage{mathtools}
\usepackage{scalerel}
\usepackage{graphicx}  
\usepackage{bm}        
\usepackage{tikz}
\usepackage{amssymb}   
\usepackage{lipsum}
\usepackage{lineno}
\usetikzlibrary{svg.path}
\usepackage{amsmath,array}
\usepackage{hyperref}
\hypersetup{
	colorlinks=true,
	linkcolor=blue,
	filecolor=magneta,      
	urlcolor=blue,
}
\DeclareCaptionJustification{justified}{\leftskip=0pt \rightskip=0pt \parfillskip=0pt plus 1fil}

\definecolor{orcidlogocol}{HTML}{A6CE39}
\tikzset{
    orcidlogo/.pic={
        \fill[orcidlogocol] svg{M256,128c0,70.7-57.3,128-128,128C57.3,256,0,198.7,0,128C0,57.3,57.3,0,128,0C198.7,0,256,57.3,256,128z};
        \fill[white] svg{M86.3,186.2H70.9V79.1h15.4v48.4V186.2z}
        svg{M108.9,79.1h41.6c39.6,0,57,28.3,57,53.6c0,27.5-21.5,53.6-56.8,53.6h-41.8V79.1z M124.3,172.4h24.5c34.9,0,42.9-26.5,42.9-39.7c0-21.5-13.7-39.7-43.7-39.7h-23.7V172.4z}
        svg{M88.7,56.8c0,5.5-4.5,10.1-10.1,10.1c-5.6,0-10.1-4.6-10.1-10.1c0-5.6,4.5-10.1,10.1-10.1C84.2,46.7,88.7,51.3,88.7,56.8z};
    }
}
\newcommand\orcidicon[1]{\href{https://orcid.org/#1}{\mbox{\scalerel*{
                \begin{tikzpicture}[yscale=-1,transform shape]
                \pic{orcidlogo};
                \end{tikzpicture}
            }{|}}}}

\begin{document}

\preprint{APS/123-QED}

\title{Constraining Primordial Black Holes via p-wave annihilation in light of CMB Spectral Distortion and 21-cm global signal}
\author{Shibsankar Si$^{\orcidicon{0009-0001-8038-976X}}$\,}
\email{shibsankar1018@gmail.com}
\affiliation{National Institute of Technology, Meghalaya, Saitsohpen, Cherrapunji, Meghalaya, 793 108, India}
\author{Pravin Kumar Natwariya$^{\orcidicon{0000-0001-9072-8430}}$\,}
\email{pvn.sps@gmail.com}
\affiliation{School of Fundamental Physics and Mathematical Sciences, Hangzhou Institute for Advanced Study, UCAS, Hangzhou, 310 024, China}
\affiliation{University of Chinese Academy of Sciences, Beijing, 100 190, China}
\author{Alekha C. Nayak$^{\orcidicon{0000-0001-6087-2490}}$\,}%
\email{alekhanayak@nitm.ac.in}
\affiliation{National Institute of Technology, Meghalaya, Saitsohpen, Cherrapunji, Meghalaya, 793 108, India}%

\date{\today}

\begin{abstract}
 Primordial black holes (PBHs) can form spike density halos through the accretion of weakly interacting massive particles (WIMPs). In these halos, the enhanced density significantly boosts the annihilation rate of WIMPs. For Majorana dark matter annihilation into light fermions, the s-wave part of the annihilation cross section is helicity-suppressed, making the p-wave contribution dominant. We study the velocity-dependent p-wave annihilation case, whose resulting energy injection can modify the thermal and ionization history of the Universe, leaving observable imprints on the cosmic microwave background (CMB) spectrum and the global 21-cm signal. From the predicted energy injection into the plasma, we derive stringent upper limits on the fraction of dark matter in form of PBHs for p-wave annihilation models, based on the observational constraints of the CMB spectral distortions ($y$-type), and from the measurement of the 21-cm absorption signal at cosmic dawn. Our results highlight that accounting for the p-wave nature of annihilation is crucial for deriving robust constraints on the PBH abundance.



\end{abstract}

\maketitle

\prlsec{\label{sec:level1} Introduction} Astronomical observations and theoretical studies suggest that around $27\%$ of the total energy budget in the Universe consists of dark matter (DM). Still, the nature of DM remains mysterious \cite{Arbey_2021, Bertone_2005, Jungman_1996, Planck:2018vyg}. DM may be composed of WIMPs, which only weakly interact with the standard model particles. Further, there is also a possibility that some fraction of total dark matter may be composed of  PBHs \cite{Planck:2018vyg, Bertone_2005, Jungman_1996, PhysRevLett.117.061101, Arcadi_2018, Fermi-LAT:2016uux, HESS:2016mib, 2016, Bird_2016, Laha:2019ssq}.

PBHs accrete mass due to their strong gravitational fields, allowing them to attract nearby matter \cite{PhysRevLett.116.041302, Mack_2007, Bertschinger:1985pd, Fillmore:1984wk,  Tiwari:2025qqx}. Several factors, including the density distribution, relative velocities between particles of the surrounding matter, and the mass of the PBHs, determine the rate and efficiency of accretion \cite{Mack_2007, Fillmore:1984wk, PhysRevLett.103.161301, Bertschinger:1985pd, PhysRevD.107.083523, Tiwari:2025qqx, Suzuki:2025kiv, Meighen-Berger:2025hrq}. Due to the accretion of WIMPs, high-density profile halos form around PBHs. Here, we consider a scenario where PBHs constitute only a fraction of DM. The remaining DM (WIMPs) are expected to form DM halos around PBHs. We are assuming these halos are spherically symmetric and follow a  density profile: $\rho(r) \propto r^{-9/4}$ \cite{Scott:2009tu, Ricotti:2009bs, Gondolo:1999ef, Clark:2016pgn, Edwards:2019tzf, Boudaud:2021irr, Tashiro:2021xnj, Kadota:2021jhg,  PhysRevLett.133.021401, Eroshenko:2016yve}. Since the annihilation rate depends on the square number density of WIMPs, it is expected to be significantly enhanced within halos. 
 The annihilation cross section can be expanded in the power of the WIMP relative velocity, $v$, as $\langle \sigma v \rangle \approx a + bv^2$, where $a$ and $bv^2$ correspond to s-wave and p-wave contributions, respectively. If the s-wave is allowed, it dominates over the p wave from freeze-out to the present epoch. The p-wave provides a non-negligible contribution at freeze-out ($\langle v^2 \rangle\approx0.1$), but it becomes negligible today due to strong velocity suppression [$\langle v^2 \rangle\approx\mathcal{O}(10^{-6})$]. For certain dark matter candidates (e.g., Majorana fermions) where s-wave is forbidden or strongly suppressed, then p-wave provides the dominant contribution even if its magnitude is small \cite{Kumar_2013, Kadota:2021jhg}. In particular, for annihilation of Majorana DM to light SM fermions with mass $m_f$ via light mediator ($\chi\chi\rightarrow f \bar{f}$), the s-wave contribution is helicity-suppressed by $m_f^2/m_\chi^2$, so the leading contribution comes from the p-wave annihilation \cite{PhysRevLett.115.231302, Bambhaniya:2016cpr, Flores:1989ru, Ciafaloni:2011sa, Kumar_2013}. Here, we show that velocity-dependent p-wave annihilation modifies the WIMP dark-matter halo profile relative to the standard s-wave case, resulting the bounds in the p-wave scenario are relaxed compared to the s-wave case.

These spiked density halos provide a novel approach to constrain the PBHs abundance. The formation of these spiked density halos has been discussed in the previous articles \cite{Eroshenko:2016yve, Boudaud:2021irr}. The observations of micro-lensing signatures also suggested the existence of the ultra-compact mini halo around PBHs \cite{Ricotti:2009bs}. In these halos, the gamma-ray emission signal is computed from the annihilation of WIMPs in Refs. \cite{Eroshenko:2016yve, Hertzberg:2020kpm, Adamek:2019gns}. 
In Ref.\cite{Kadota:2022cij}, the authors study the synchrotron radio emission from WIMP annihilation within these halos.
In this letter, we want to explore the p-wave annihilation Scenario of dark matter particles in these halos and calculate an upper bound using the global 21-cm signal and CMB spectral distortion observations. After matter-radiation equality, secondary infall drives the formation of spike density around PBH, which enhances the annihilation rate \cite{Hertzberg:2020kpm}. Energy released from WIMP annihilation between recombination and matter-radiation equality induces the CMB $y$-type distortion, while in post-recombination annihilation affects the 21-cm signal.

The CMB spectral distortion is a successful approach to understanding the evolution of the Universe. The CMB follows the blackbody radiation spectrum \cite{ 1991ApJ...371L...1S, Mather:1993ij}. If energy is injected into plasma at a redshift $z>2\times10^6$, it is rapidly redistributed into photons through the Compton scattering, while photon number density is maintained by Bremsstrahlung and double Compton scattering to preserve the blackbody spectrum \cite{PhysRevLett.115.071304, khatri2012beyond, PhysRevD.111.043002, Natwariya:2025ftu}. However, below $z<2\times10^6$, Bremsstrahlung and Double Compton scattering become inefficient. The energy injection during $5\times10^4\leq z \leq 2\times10^6$ can deviate the CMB spectrum from a perfect blackbody, producing the Bose-Einstein spectrum with a non-zero chemical potential $\mu$, called a $\mu$-distortion \cite{khatri2012beyond, Kunze:2013uja, Si:2024qgu}. Below the redshift of $ z \leq 5\times10^4$, Compton scattering becomes ineffective. Any energy injection into plasma no longer produces the Bose-Einstein distribution, resulting in the $y$-type of spectral distortion \cite{Si:2024qgu}. Any energy injection into plasma produces $y$-type distortion between $10^3 \lesssim z\lesssim 5\times10^4 $, where photons gain energy by interacting with electrons via the inverse Compton scattering, and low-energy CMB photons are up-scattered by high-energy electrons \cite{Sunyaev:1970er}. In this work, we consider the energy injection from WIMP annihilation during the epochs of recombination and matter–radiation equality, which leads to the generation of $y$-type distortions in the CMB spectrum.
The parameter $y$ can be rewritten as follows \cite{Si:2024qgu}:
      \begin{eqnarray}\label{eq: y parameter}
           y&=&\frac{1}{12}\frac{\delta \rho_\gamma}{\rho_\gamma}=\frac{1}{12}\int \Big[\Big(\frac{dE_{\rm inj}}{dVdz}\Big)\Big/\rho_\gamma\Big]dz
      \end{eqnarray}
 Where $({dE_{inj}}/{dVdz})=-({dE_{inj}}/{dVdt})/[H(1+z)]$ is the injected energy per unit time per unit volume and $\rho_\gamma\equiv\rho_\gamma(z)$ is the energy density of the CMB photons at the corresponding redshift. The Far Infrared Absolute Spectrophotometer (FIRAS) and the projected sensitivity of Primordial Inflation Explorer (PIXIE) put an upper limit on the CMB spectral distortions \cite{Fixsen:1996nj, Kogut:2011xw}. The FIRAS experiment has the constraint on the $y$ parameter equal to $y\approx 1.5\times10^{-5}$ \cite{Fixsen:1996nj, Mather:1993ij}, where PIXIE will able to detect the $y$ parameter at the level of $y\approx10^{-8}$ at $5\sigma$ \cite{Kogut:2011xw}.

After recombination (at $z\approx 10^3$), the baryonic matter was primarily composed of neutral hydrogen and a small fraction of helium. Within the neutral hydrogen atoms, the ground state splits into two hyperfine levels due to the spin-spin interaction between the electron and proton: the singlet state ($F=0$) and the triplet state ($F=1$). The relative population densities of these states are governed by the relation $n_1/n_0 = (g_1/g_0) \exp(-T_*/T_s)$, where $g_1=3$ and $g_0=1$ are the degeneracies of the triplet and singlet states, respectively. Here, $T_* = 68,\mathrm{mK}$ is the temperature equivalent to the photon energy released in the transition between these two levels, which corresponds to a rest wavelength of 21 cm (a frequency of approximately 1420 MHz). The $T_s$ is the spin temperature, which determines the relative population density of these states\cite{Si:2025vsj, DAmico:2018sxd}. The difference between the spin temperature $T_s$ and the background temperature $(T_{\gamma})$, defines the brightness temperature $(\delta T_b)$, which is given by \cite{ PhysRevLett.121.011103, PhysRevLett.118.151301, PhysRevLett.121.121301, PhysRevLett.133.131001, PhysRevLett.121.011101, Mitridate:2018iag, Cyr:2023iwu, PhysRevD.110.123506}
\begin{equation}\label{eq: T21}
    \delta T_b \approx 27x_{\rm HI}\left(1-\frac{T_\gamma}{T_s}\right)\left(\frac{1+z}{10}\right)^{0.5} \left(\frac{0.15}{\Omega_m}\right)^{0.5}\left(\frac{\Omega_bh}{0.023}\right)\mathrm{mK}\, ,
\end{equation}
where $x_{\rm HI}=\frac{n_{\rm HI}}{n_H}$ is the neutral hydrogen fraction, $n_{HI}$ is the neutral hydrogen number density, and $n_H$ is the total hydrogen number density. $\Omega_m$ and $\Omega_b$ are the dimensionless density parameters of matter and baryons, respectively. The spin temperature is determined by scattering with CMB photons, collisional coupling of the baryonic gas ($T_b$), and interactions with $\rm Ly\alpha$ photons from the first stars. It is given by $T_s^{-1} = \frac{T_{\gamma}^{-1}+x_c{T_{b}^{-1}+x_\alpha T_\alpha^{-1}}}{1+x_c+x_\alpha}$, where $T_\alpha$ is the color temperature, which is closely coupled to $T_b$. Here, $x_c$ and $x_\alpha$ are the collisional and $\rm Ly\alpha$ coupling coefficients, respectively. The absorption signal we can observe in the 21 cm signal when $T_s< T_\gamma$. The injected energy from WIMP annihilation increases the baryon temperature $T_b$, leading to a reduction in the depth of the 21-cm brightness temperature $(\delta T_b)$.

\prlsec{\label{II} The Properties of spike density halo} 
The size of the spike density halos can be defined by the turnaround radius $r_{ta} \approx (R_{S}\, t^{2}_{ta})^{1/3}$ \cite{Kadota:2021jhg}, where the gravitational attraction of a PBH causes a DM mass shell to decouple from the Hubble flow. Where $R_{S}=2\,G\,M_{\rm PBH}$ is the Schwarzchild radius,  the turnaround time---$ t_{ta}$ is when WIMP starts falling towards PBHs at radius $r_{ta}$. During the matter-radiation equality period, the WIMP halo density profile can be approximated as \cite{Kadota:2021jhg}
\begin{equation}\label{eq: spike density}
    \rho_{sp}(r) \approx {\rho_{\chi}(z_{eq})} \left(\frac{r}{r_{ta}(z_{eq})}\right)^{-9/4} \,  for \,\,\, r<r_{ta}(z_{eq})
\end{equation}
where, $\rho_{\chi}(z_{eq})$ is the background DM density at matter–radiation equality. After matter–radiation equality, dark matter halos can continue to grow through a secondary infall mechanism onto the halo described by Eq.\eqref{eq: spike density} \cite{Gines:2022qzy}. Even during the matter domination epoch ($z_{\rm eq}$), the spike slope in the inner halo region ($r < r_{ta}(z_{eq})$) remains stable, while the infalling matter develops an NFW-like profile at large radii \cite{Navarro:1995iw}. Therefore, we considered that the density profile in Eq.\eqref{eq: spike density} remains valid for $z<z_{eq}$. DM annihilation may cause the innermost center region to be shallower than the outer spike. The resulting steep density profile around a PBH is modeled as \cite{Kadota:2021jhg} 
\begin{equation}\label{eq: density}
\rho(r) = 
\begin{cases}
0 & \text{for }r < 2R_s \\
\rho_{\rm core} & \text{for } 2R_s <r < r_{\rm core} \\
{\rho_{\chi}(z_{eq})} \left(\frac{r}{r_{ta}(z_{eq})}\right)^{-9/4} & \text{for } r_{\rm core} < r < r_{ta}(z_{eq}). \\
\end{cases}
\end{equation}
For $r < 2 R_{S}$, the halo density disappears because the WIMP particles are absorbed by PBH. Here, $r_{core}$ is the core radius at which $\rho_{\rm core}(r_{\rm core})=\rho_{sp}(r_{\rm core})$. In the presence of annihilation, the inner central density profiles can be given as \cite{Kadota:2022cij}
\begin{equation}\label{eq: core density}
    \rho_{\rm core} \approx \frac{m_{\chi}}{\langle \sigma v \rangle (t - t_{i})},
\end{equation}
where $m_{\chi}$ is the WIMP mass, $t_{i}$ is the halo formation time, we consider $t_{i} = t_{eq}$ is the time at matter radiation equality; consequently, the amplitude of the inner core density $(\rho_{\rm core})$ decreases with time. The thermally averaged WIMP annihilation cross section is $\langle \sigma v \rangle\approx \langle\sigma v \rangle_s + \langle\sigma v \rangle_p$, where the p-wave term scales as $\langle \sigma v\rangle_p \propto v^2$. For Majorana fermions annihilating into light SM fermions, the s-wave contribution $\langle\sigma v \rangle_s$ is helicity suppressed; therefore, the leading contribution arises from the p-wave term $\langle \sigma v\rangle_p$.
Considering the local relative velocity distribution as approximately the Maxwell-Boltzmann distribution determined via the virial theorem \cite{Kadota:2021jhg}. Under this consideration, the thermal average p-wave annihilation cross section is given by \cite{Kadota:2021jhg}
 \begin{eqnarray}\label{eq: cross section}
     \langle \sigma v \rangle_p &= &\langle \sigma v \rangle_{o}^p \ (v/v_{o})^2 ,
 \end{eqnarray}
where, $v_{o}\approx 0.3$ is the WIMP velocity at freeze-out  \cite{Kadota:2021jhg}, and $\langle \sigma v \rangle_{o}^p \approx 6\times10^{-26}$ $\rm cm^3 s^{-1}$ is the p wave annihilation cross section at freeze-out \cite{Kadota:2021jhg}. 

\begin{figure}
\centering
\includegraphics[width=\linewidth]{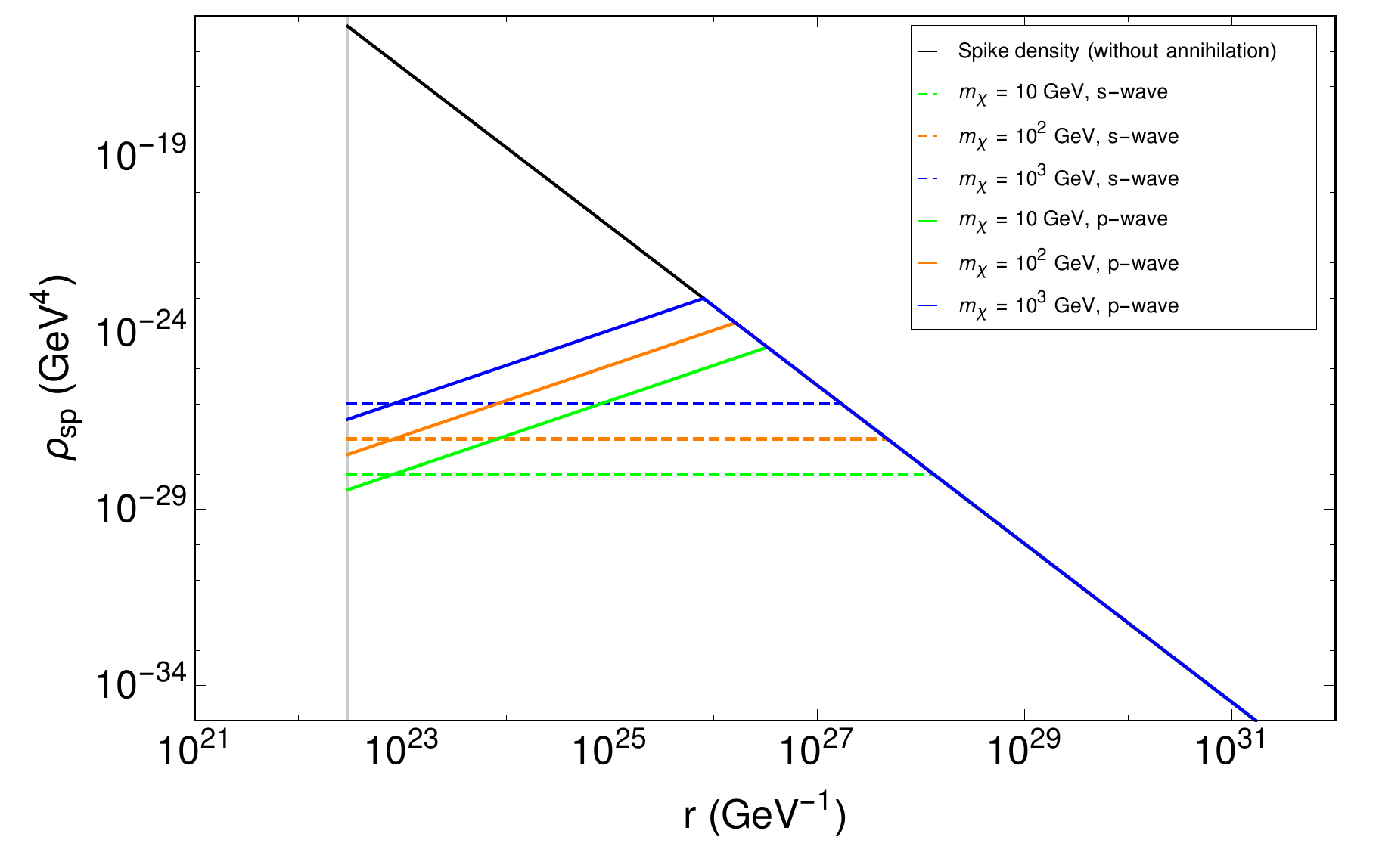}
\caption{The density profile of WIMP annihilation as a function of radial distance. The black solid line corresponds to the density profile before WIMP annihilation. In the solid and dashed Coloured lines denote the effect of p-wave and s-wave WIMP annihilation, respectively, at redshift $z=1000$ for $\rm M_{PBH}=10^3\, M \odot $. The gray solid line indicates $r=2R_s$; for $r < 2 R_{S}$, the WIMPs are absorbed by PBH. }
\label{plot: density}
\end{figure}

The WIMP velocity in the halo around PBH is given by, $v=\sqrt{G\, M_{\rm PBH}/r}$, with $G$ is the gravitational constant, $M_{\rm PBH}$ is the PBH mass, and $r$ is the radial distance from the halo center \cite{Gines:2022qzy}. For the PBH with solar mass, the WIMP velocity inside the halo at radius $r\approx2R_s$ is $v\sim 0.5$, where the outside of the halo, at the turnaround radius $t_{ta}(z_{eq})$, the velocity drops to $\mathcal{O}(10^{-6})$. Consequently, p-wave annihilation can be significant in the inner halo, while it is strongly suppressed at large radii. As a result, for s-wave annihilation scenarios, the inner core region ($r<r_{\rm core}$) may form a flat plateau due to the constant annihilation cross section (see figure \ref{plot: density}). In contrast, in p-wave annihilation scenarios, the plateau is not flat, as the annihilation rate depends explicitly on velocity (radius) [see Eqs. \eqref{eq: core density} and \eqref{eq: cross section}].

A large density can significantly affect the annihilation because the DM annihilation rate is proportional to the square of the WIMP density. The annihilation rate of the WIMP in these halos is given by
\begin{eqnarray}\label{eq: annihilation rate}
     \Gamma_{\rm WIMP} =\int n_{\chi}^{2} \langle \sigma v \rangle dV =\frac{4 \pi }{m_{\chi}^{2}}\int_{2R_s}^{r_{ta}(z_{eq})} r^{2} \rho(r)^{2} \langle \sigma v \rangle\, dr,\,
\end{eqnarray}
where $m_{\chi}$ and $n_\chi$ are the mass and number density of the WIMP in the spiked density halos. The $r_{ta}(z_{eq})$ is the turnaround radius at matter radiation equality. 

The energy injection from WIMP annihilation per unit time per unit volume can be written as,
\begin{eqnarray}\label{eq: energy injection}
    \frac{dE_{\rm inj}}{dVdt}&=&n_{\rm PBH}\,m_\chi\,\Gamma_{\rm WIMP}\nonumber\\
    &= &f_{\rm PBH}\,\Omega_{\rm DM}\,\rho_c\,\Gamma_{\rm WIMP}\,\frac{m_\chi}{M_{\rm PBH}}\,(1+z)^3\,,
\end{eqnarray}
here $n_{\rm PBH}=\rho_{\rm PBH}/M_{\rm PBH}$ is the number density of PBHs, where $\rho_{\rm PBH}=\rho_c\,\Omega_{\rm PBH}\, (1+z)^3$, $\rho_c$ is the present-day critical density of the Universe. The abundance of PBHs $\Omega_{\rm PBH}=f_{\rm PBH}\,\Omega_{\rm DM}$, where $f_{\rm PBH}$ is the fraction of DM in the form of PBHs and $\Omega_{\rm DM}$ is the DM density parameter.   

Depending on the photon energy and the plasma state, photons can have a sizable mean free path and may not be absorbed instantaneously. In such cases, the energy deposition rate into the plasma can be expressed as,
\begin{align}\label{eq: energy deposition}
    \frac{dE_{\rm dep}}{dVdt}=f_{\rm abs}(z,m_{\chi})\ \frac{dE_{\rm inj}}{dVdt}\,
\end{align}
here, $f_{\rm abs}(z,m_{\chi})$ represents the fraction of energy deposition from the injected energy \cite{PhysRevD.87.123513, PhysRevD.93.023527, PhysRevD.93.023521}.

\prlsec{The temperature evolution of the baryon }
The thermal evolution of the baryon in the presence of energy injection from WIMP annihilation is given by \cite{Peebles:1968ja, Seager:1999bc, Seager:1999km},
\begin{alignat}{2}\label{eq: Tb}
    (1+z)H(z)\frac{dT_{b}}{dz} = 2H(z)T_{b}-{\Gamma_C}(T_{\gamma}-T_{b})\notag \\
    - \frac{2}{3} \frac{(1 + 2x_e)}{3N_{\text{tot}}} 
     \frac{dE_{\rm dep}}{dV\,dt},
\end{alignat}
where, $H(z)$ is the the Hubble parameter, $N_{\rm tot}=n_H(1+f_{\rm He}+x_e)$ is the total baryon number density. The ionization fraction $x_e$ is defined as ${n_e}/{n_H}$, and the helium fraction is $f_{\rm He}={n_{\rm He}}/{n_H}$, where $n_H$, $n_e$, and $n_{He}$ are hydrogen, electron, and helium number density, respectively. Further, $\Gamma _C= \frac{8\sigma_Ta_rT_\gamma^4x_e}{3m_e(1+f_{\rm He}+x_e)}$ is the Compton scattering rate, where $\sigma_T$ is the Thomson scattering cross section and the radiation constant, and  $a_r$ is the Stefan-Boltzmann radiation constant. 

The evolution of the ionization fraction, $x_e$, in the presence of WIMPs annihilation is given by \cite{Peebles:1968ja, Seager:1999bc, Seager:1999km},
\begin{alignat}{2} \label{eq: Xe}
      (1+z)H(z) \frac{dx_e}{dz} = {\cal C}\left[n_HA_Bx_e^2 - 4(1 - x_e)B_B e^{{-3E_{0}}/{4T_{\gamma}}}\right] \notag \\
   -\frac{1 - x_e}{N^{\text{tot}}}\left( \frac{\mathcal{C}}{E_0} + \frac{1 - \mathcal{C}}{E_\alpha} \right) 
    \frac{dE_{\rm dep}}{dV\,dt}\,, 
\end{alignat} 
where $E_{0}= 13.6\,$eV is the ground state energy and $E_\alpha\approx {3}/{4}\,E_0$ is the energy of $\rm Ly\alpha$ photon \cite{Ali-Haimoud:2010hou, Ali-Haimoud:2010tlj}, $\cal C$ is the Peebles coefficient \cite{Peebles:1968ja, DAmico:2018sxd}. $B_B(T_\gamma)$ is the case-B photo-ionization rate and $A_B(T_{gas})$ is the case-B recombination rate as \cite{Peebles:1968ja, Seager:1999bc, Seager:1999km, Bhatt:2019lwt}.

\begin{figure}
\centering
\includegraphics[width=\linewidth]{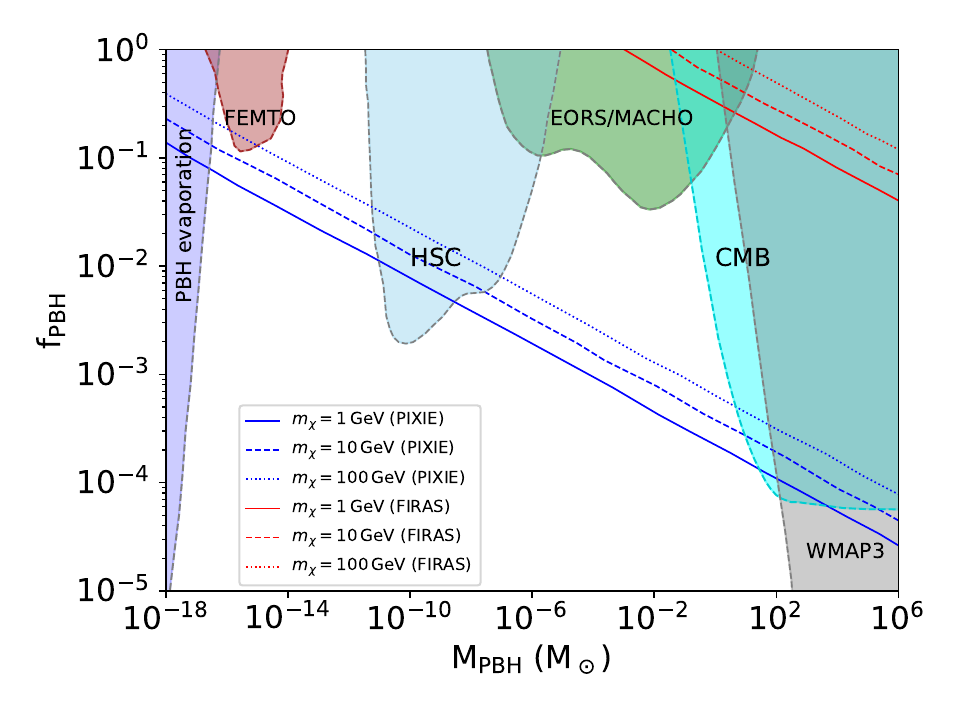}
\caption{Upper limits on the fraction of DM in the form of PBHs, $f_{\rm PBH}= \Omega_{\rm PBH}/\Omega_{\chi}$ as a function of PBH mass, $M_{\rm PBH}$. The bounds have been obtained using FIRAS and PIXIE limits on $y$-type of spectral distortion. In this plot, Solid, dashed, and dotted lines represent the upper bound for p-wave annihilation with WIMP mass $m_\chi=1\,\rm GeV,\,10\,\rm GeV,$ and $100\, \rm GeV$, respectively. Our upper bound compared with other observational constraints, from extragalactic gamma-rays observation of PBH evaporation(blue region), femto-lensing (brown region) \cite{Barnacka:2012bm}, microlensing HSC/Subaru data (sky-blue region) \cite{Niikura:2017zjd},  microlensing EROS/MACHO(green region) \cite{Niikura:2017zjd}, X-ray emitted by gas accretion CMB (cyan region), and WMAP3(gray region) \cite{Ricotti:2007au}.}
\label{plot: distortion}
\end{figure}

\prlsec{\label{result}Results and Discussion} We derive upper limits on the fraction of dark matter in the form of primordial black holes, $f_{\rm PBH} \equiv \Omega_{\rm PBH}/\Omega_{\rm DM}$, assuming the remaining dark matter consists of WIMPs undergoing velocity-suppressed (p-wave) annihilation within density spikes formed around PBHs. The constraints are obtained from current COBE/FIRAS limits on y-type CMB spectral distortions ($y < 1.5 \times 10^{-5}$), the projected PIXIE sensitivity ($y \sim 10^{-8}$ at 5$\sigma$), and the requirement that the deep 21-cm absorption trough observed by EDGES at cosmic dawn ($z \simeq 17$) is not erased--- i.e. $T_b<T_{\rm CMB}$, by excess heating or ionization. 


Figure \ref{plot: density} illustrates the radial density profiles of WIMP dark matter in the spike halos at $z = 1000$ for $M_{\rm PBH} = 10^{3}\,M_{\odot}$. For s-wave annihilation the inner region forms a flat plateau because $\langle \sigma v \rangle$ is velocity-independent. In contrast, p-wave annihilation ($\langle \sigma v \rangle \propto v^{2}$) produces a softer, radially dependent core because the local velocity $v(r) \propto r^{-1/2}$ decreases toward the PBH, suppressing annihilation in the innermost regions and allowing higher central densities (see Eqs. \ref{eq: cross section} and \ref{eq: core density}). This velocity suppression is the key physical difference that relaxes the p-wave bounds relative to the s-wave case.


The resulting annihilation rate and energy injection into plasma are computed using Eqs. (\ref{eq: annihilation rate}-\ref{eq: energy deposition}). The injected energy modifies the thermal and ionization history of the Universe through Eqs. \eqref{eq: Tb} and \eqref{eq: Xe}, which are solved numerically with initial conditions $T_b = 2758$ K and $x_e = 0.05725$ at $z = 1010$.

\begin{figure}
\centering
\includegraphics[width=\linewidth]{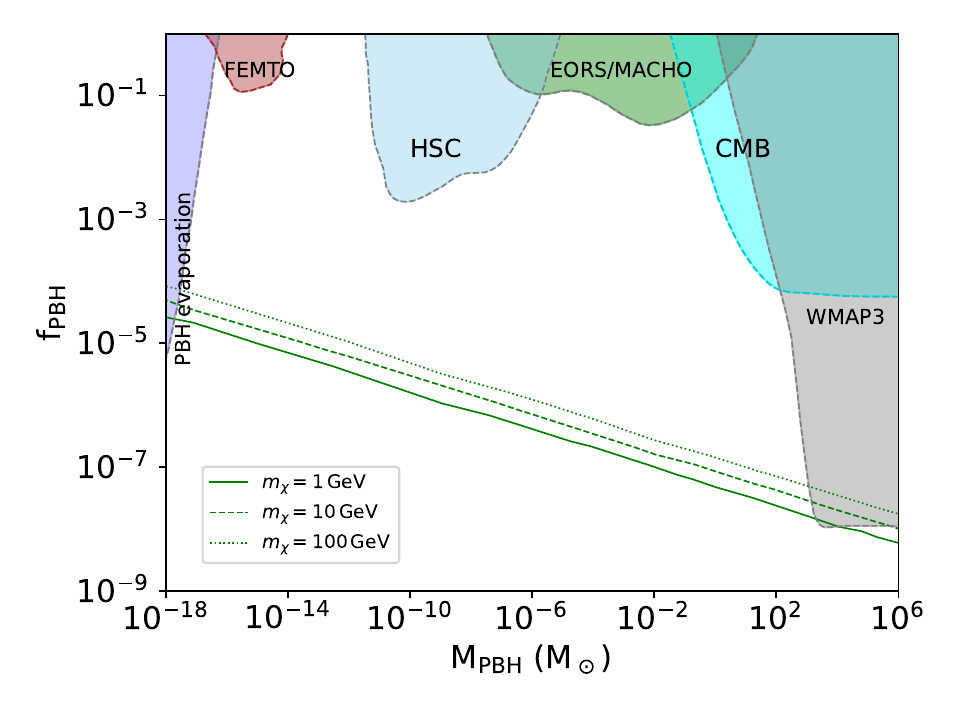}
\caption{Upper limits on the fraction of DM in the form of PBHs, $f_{\rm PBH}$ as a function of PBH mass, $M_{\rm PBH}$. We obtained bounds such that the WIMP annihilation will erase the amplitude of the 21 cm brightness temperature ($\delta T_b=0$) at $z=17$. Here, Solid, dashed, and dotted lines represent the upper bound for p-wave annihilation with DM mass $m_\chi=1\,\rm GeV,\,10\,\rm GeV,$ and $100\, \rm GeV$, respectively.}
\label{plot: 21cm}
\end{figure}

Figure \ref{plot: distortion} shows the upper bounds on $f_{\rm PBH}(M_{\rm PBH})$ derived from y-type CMB spectral distortions. The red curves correspond to the current FIRAS limit, while the blue curves obtained using the forecasted PIXIE sensitivity. Solid, dashed, and dotted lines denote WIMP masses $m_\chi = 1,\, 10$ and $100$ GeV, respectively. For p-wave annihilation, the bound scales approximately as $f_{\rm PBH} \propto M_{\rm PBH}^{-2/13}$ (derived analytically in the Appendix of Ref. [27] and confirmed here). This mild mass dependence arises because the p-wave velocity suppression reduces the annihilation efficiency in denser, lower-velocity regions near the PBH, allowing a larger PBH abundance for a fixed energy-injection budget.

Using the current FIRAS bound, p-wave annihilation yields $f_{\rm PBH} \lesssim 1.0$ – $3.99\times 10^{-2}$ over the PBH mass range $1.09\times 10^{-3}\,M_\odot$ to $10^6 \,M_\odot$ for WIMPs mass $m_\chi=1\, {\rm GeV}$. With PIXIE’s projected sensitivity, the limits strengthen to $f_{\rm PBH} \lesssim 1.3 \times 10^{-1}$ – $2.6 \times 10^{-5}$ across the broader range $10^{-18}\,M_\odot$ to $10^{6}\,M_\odot$. Bounds weaken with increasing $m_\chi$ because the WIMP number density $n_\chi \propto m_\chi^{-1}$ decreases, reducing the annihilation rate. Consequently, for a fixed energy injection, the allowed value of $f_{\rm PBH}$ increases.


Figure \ref{plot: 21cm} presents constraints from the global 21-cm signal, requiring that excess WIMP annihilation does not erase the absorption trough ($\delta T_b \lesssim 0$ mK at $z \simeq 17$) by overheating the gas or ionization. These limits are significantly stronger than those from CMB distortions, reaching $f_{\rm PBH} \lesssim 2.6 \times 10^{-5}$–$5.6 \times 10^{-9}$ for $m_\chi = 1$ GeV across the same wide PBH mass range. The green curves again correspond to $m_\chi = 1, 10,$ and $100$ GeV (solid, dashed, dotted).


For the s-wave annihilation, the bound on $f_{\rm PBH}$ is independent of $\rm M_{PBH}$ (see in the Appendix Fig. \ref{plot: contour_swave}). As the PBH mass increases, the WIMP density in the halo surrounding each PBH also increases. Since the annihilation rate depends on the square of the WIMP density, this leads to an enhanced annihilation rate around PBHs. As a result, the total WIMP annihilation rate scales linearly with the PBH mass, as $\Gamma_{\rm WIMP}\propto \rm M_{PBH}$. On the other hand, for a fixed PBH abundance, the number density of PBHs decreases inversely with their mass, as $\rm n_{PBH}\propto1/M_{PBH}$. As a result, the energy injection rate is independent of PBH mass [$(dE_{inj}/dVdt)\propto n_{\rm PBH}\,\Gamma_{\rm WIMP}$]. Consequently, for a fixed CMB spectral distortion and 21-cm brightness temperature, the corresponding upper bounds on $ f_{\rm PBH}$ also become independent of PBH mass. 
In contrast, in the p-wave case, the bounds on $f_{\rm PBH}$ are relaxed. It explicitly depends on the PBH mass, as discussed above . The bound on $ f_{\rm PBH}$ also depends on the WIMP mass $m_\chi$. The bound becomes weaker for larger WIMP masses, since an increase in mass reduces the number density of WIMPs; as a result, the annihilation rate will be decreased and the corresponding bounds on $ f_{\rm PBH}$ become relaxed. 

In both figures (\ref{plot: distortion}) and (\ref{plot: 21cm}), our p-wave bounds are compared with existing constraints: extragalactic gamma-ray observation of PBH evaporation (blue shaded region), femto-lensing of gamma-ray bursts (brown shaded region) \cite{Barnacka:2012bm}, microlensing surveys (HSC/Subaru-- sky-blue shaded region, EROS/MACHO-- green shaded region) \cite{Niikura:2017zjd}, and the CMB distortions and anisotropies from X-ray emitted by gas accretion using FIRAS (cyan region), and WMAP3 (gray region) \cite{Ricotti:2007au}.  Over a wide range of asteroid-to-solar-mass PBHs, the 21-cm signal provides most stringent limits on the fraction of the dark matter in the forms of primordial black holes, $f_{\rm PBH}$.

\prlsec{Conclusions} This work explores the allowed PBH abundance for p-wave annihilation in light of CMB spectral distortion and 21-cm global signal at cosmic dawn. So far, we have discussed the DM annihilation into light fermions; the s-wave is helicity suppressed by a factor $m_f^2/m_\chi^2$, so the p-wave will be dominant. The WIMP, which is gravitationally accreted onto PBH to form spike density halos with a density profile $\rho(r) \propto r^{-9/4}$. Since the annihilation rate of WIMPs is proportional to the square of their number density, it is significantly modified within these halos. The injected energy into the plasma from WIMP annihilation can affect the thermal and ionization history of the early Universe, leading to CMB spectral distortion and affecting the 21-cm global signal. We derive the bound using the FIRAS and PIXIE limits on the $y$ type of CMB spectral distortion, as well as from the global 21-cm absorption signal at redshift 17, such that the absorption amplitude remains unchanged compared to the standard $\Lambda$CDM model in the presence of WIMP annihilation. We also derive bounds at $z=15$ and $z=20$; however, no significant deviation from the bound obtained at $z=17$ (see in the Appendix Fig. \ref{plot: 21cm_z15_17_20}). The resulting bounds on the allowed PBH fraction are correspondingly relaxed for the velocity-suppressed annihilation cross section (p-wave), compared to the velocity-independent (s-wave) case. For s-wave annihilation, the bound is independent of PBH mass; in contrast, for p-wave annihilation, the bound explicitly depends on PBH mass $\sim f_{\rm PBH}\propto M_{\rm PBH}^{-2/13}$. Furthermore, the constraints also depend on the WIMP mass; they become relaxed for a higher WIMP mass.

\bibliography{main.bib}

\clearpage
\onecolumngrid

\setcounter{figure}{0}
\vspace{1 cm}

\prlsec{Appendix}

\begin{figure*}[htbp]
    \begin{center}
        \subfloat[] {\includegraphics[width=3.5in,height=2.5in]{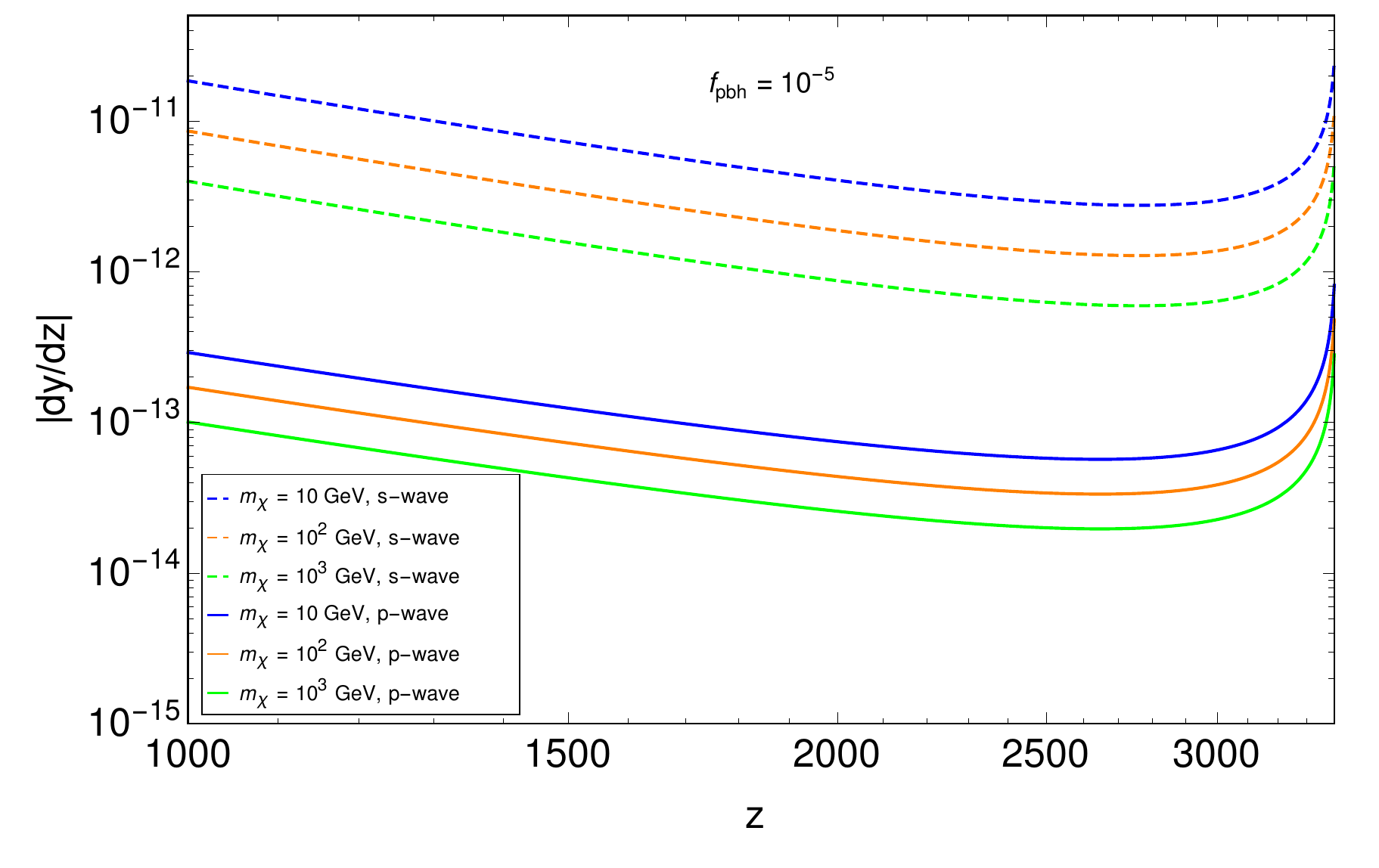}\label{dydz_f-5}}
        \subfloat[] {\includegraphics[width=3.5in,height=2.5in]{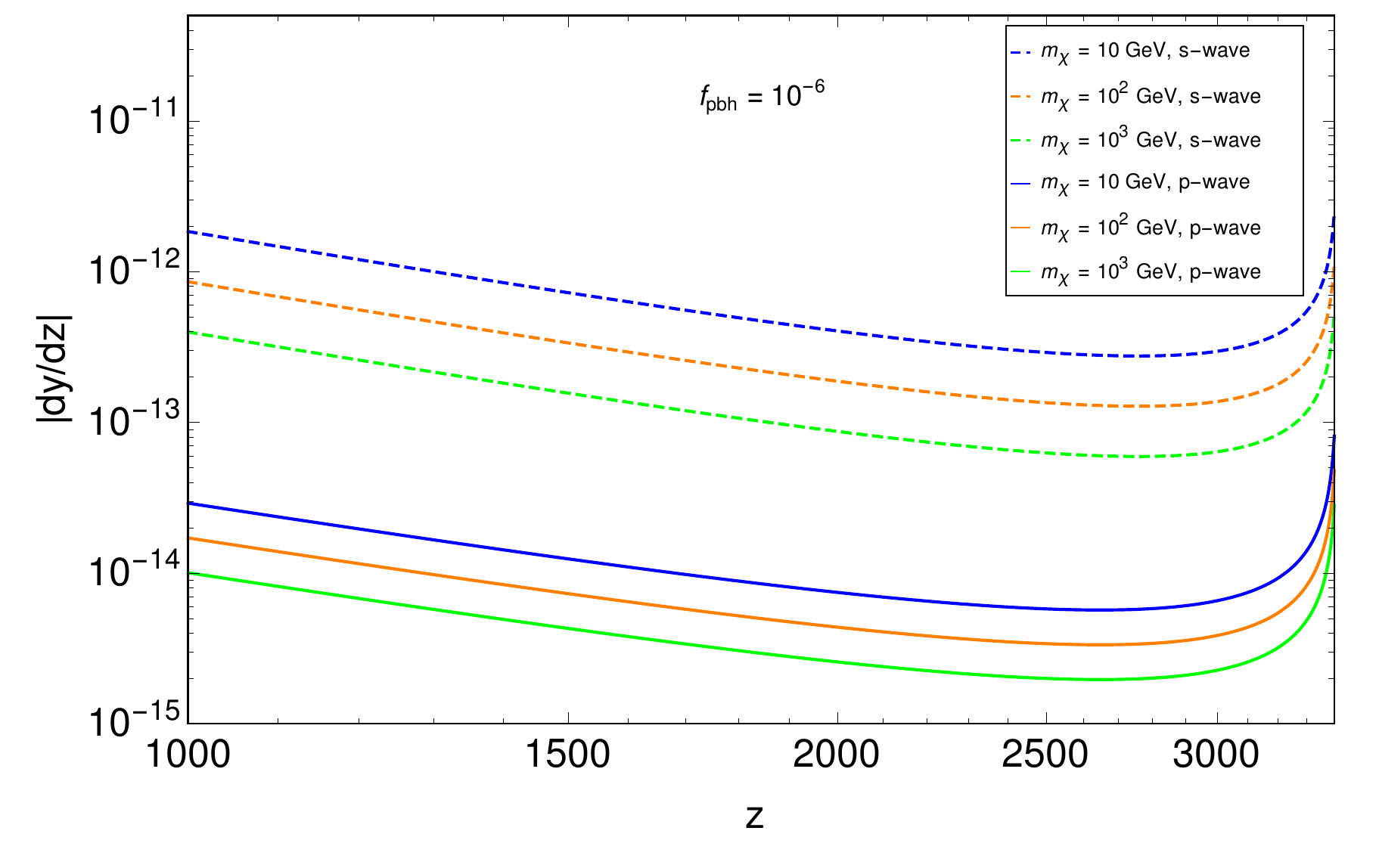}\label{dydz_f-6}}
        
    \end{center}
    \caption{Evolution of CMB spectral distortions, $|dy/dz|$, as a function of redshift z for $\rm M_{PBH}=10^3\, M \odot $. In both figures, the solid and dashed Coloured lines denote the CMB spectral distortion for p-wave and s-wave WIMP annihilation, respectively.}
    \label{plot: Tb_plot}
\end{figure*}

 \begin{figure*}[htbp]
    \begin{center}
        \subfloat[] {\includegraphics[width=3.5in,height=2.5in]{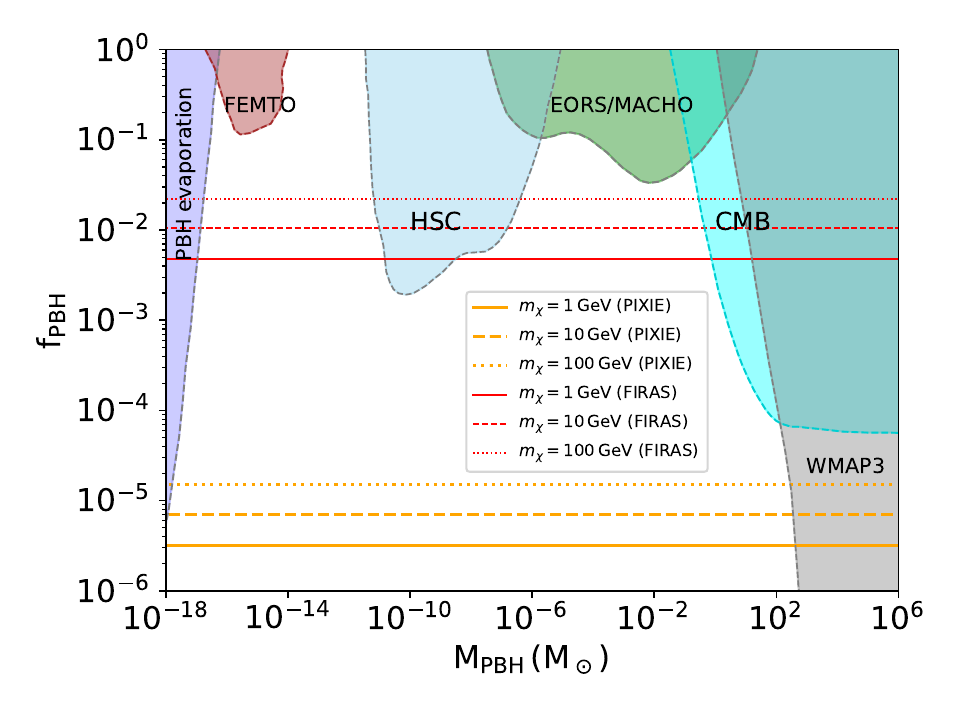}\label{s_CMB}}
        \subfloat[] {\includegraphics[width=3.5in,height=2.5in]{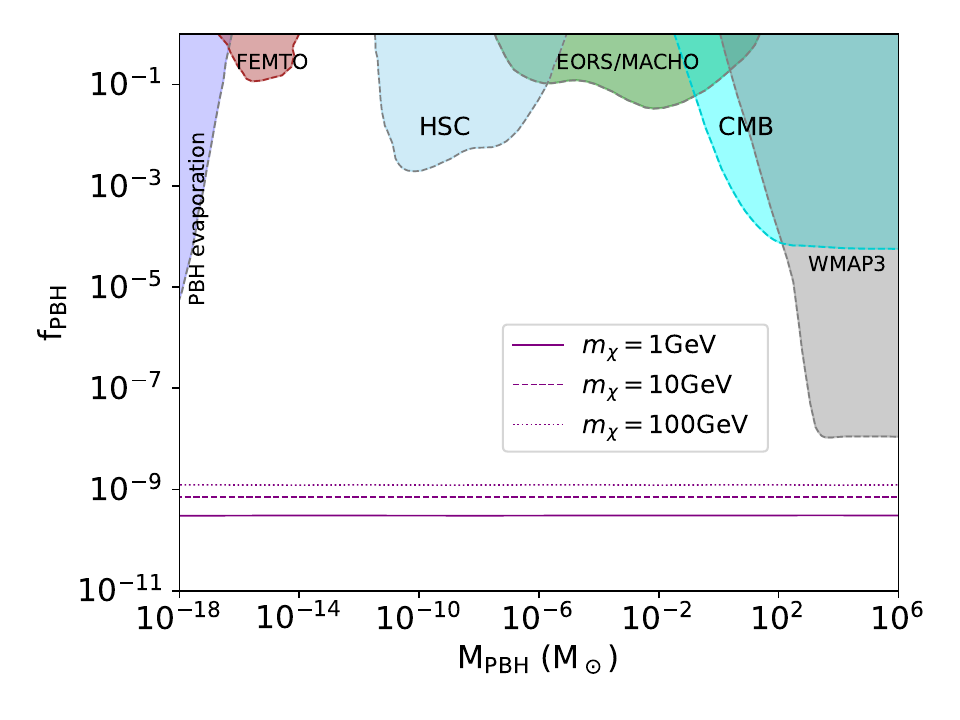}\label{s_21cm}}
        
    \end{center}
    \caption{Upper limits on the fraction of DM in the form of PBHs, $f_{\rm PBH}= \Omega_{\rm PBH}/\Omega_{\chi}$ as a function of PBH mass, $M_{\rm PBH}$. In both plots, Solid, dashed, and dotted lines represent the upper bound for the s-wave case with thermally averaged WIMP annihilation cross section $\langle \sigma v \rangle_s\approx 3\times10^{-26}$ $\rm cm^3 s^{-1}$ for WIMP mass $m_\chi=1\,\rm GeV,\,10\,\rm GeV,$ and $100\, \rm GeV$, respectively. In the left panel, the bounds have been obtained using FIRAS and PIXIE limits on $y$-type of spectral distortion, while in the right panel, the bounds are obtained by requiring ($\delta T_b=0$) at $z=17$ due to WIMP annihilation.}
    \label{plot: contour_swave}
\end{figure*}
\begin{figure}[htbp]
\centering
\includegraphics[width=3.5in,height=2.5in]{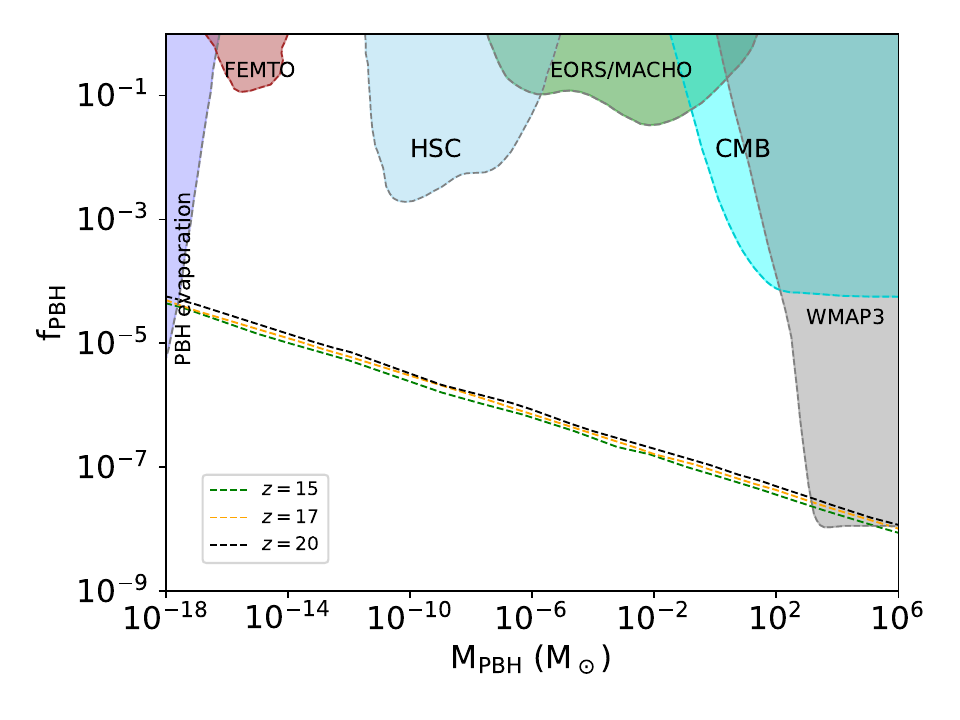}
\caption{Upper limits on the fraction of DM in the form of PBHs, $f_{\rm PBH}$ as a function of PBH mass, $M_{\rm PBH}$. We obtained bounds such that the WIMP annihilation will erase the amplitude of the 21 cm brightness temperature ($\delta T_b=0$) with DM mass $m_\chi=10\,\rm GeV$. Here, green, orange, and black dashed lines represent the upper bound for p-wave annihilation at redshifts $z=15,~17$, and $20$, respectively.}
\label{plot: 21cm_z15_17_20}
\end{figure}


\end{document}